\documentclass[conference]{IEEEtran}
\IEEEoverridecommandlockouts
\usepackage{cite}
\usepackage{amsmath,amssymb,amsfonts}
\usepackage{algorithmic}
\usepackage{graphicx}
\usepackage{textcomp}
\usepackage{threeparttable}
\usepackage{optidef}
\usepackage{xcolor}
\def\BibTeX{{\rm B\kern-.05em{\sc i\kern-.025em b}\kern-.08em
    T\kern-.1667em\lower.7ex\hbox{E}\kern-.125emX}}
\begin{document}

\title{A Unified Approach for Optimal Cruise Airspeed with Variable Cost Index for Fuel-powered and All-electric Aircraft\\
}

\author{\IEEEauthorblockN{Lucas Souza e Silva, Ali Akgunduz and Luis Rodrigues}
\IEEEauthorblockA{\textit{Department of Electrical and Computer Engineering} \\
\textit{Concordia University}\\
Montreal, Canada 
}
}
\maketitle

\begin{abstract}
This paper proposes for the first time a unified optimal approach to solve a direct operating cost (DOC) minimization problem where the cost index (CI) is time-varying. More specifically, the coefficient  CI is modeled as a time-varying parameter commanded by Air Traffic Control (ATC). The proposed unified approach relies on the solution of an optimal control problem both for fuel-powered and all-electric aircraft. Furthermore, this paper demonstrates how a variable CI affects the solution of the optimization problem as it presents the equations that allow the computation of optimal constant cruise airspeed and flight time in response to step changes in the CI value. 
The proposed methodology is validated by a simulated flight scenario.
In this scenario the inputs from the ATC are received during flight and the aircraft is required to adjust its optimal airspeed, flight time, and total energy consumption to comply with the operational restrictions imposed by the ATC.
The optimal values of airspeed, flight time and energy consumption are computed for both a fuel-powered and an all-electric aircraft, thus enabling applications of the proposed approach to future air mobility all-electric vehicles.
\end{abstract}

\begin{IEEEkeywords}
direct operating cost, flight management system, air traffic control, optimal control, fuel-powered aircraft, all-electric aircraft.
\end{IEEEkeywords}

\section{Introduction}\label{sec1}
The demand for domestic and international travel is steadily increasing. According to a recent report from the International Air Transport Association (IATA), passenger traffic in 2023 reached 94.1\% of pre-pandemic (2019) levels and it is estimated to fully recover the normal growth pattern in 2024 \cite{b1}. Airbus projects an annual demand increase in passenger traffic by 3.6\% over the next 20 years \cite{b2}, while Boeing forecasts, for the same period, a demand for 42000 new airplanes worldwide \cite{b3}. The vast majority of the aircraft fleet presently operating around the globe and that will meet the near-future demand is powered by fossil-based fuels \cite{b4}. This poses its challenges to modern aviation, such as keeping the market profitable while adjusting the fuel consumption to levels that conform to the new environmental guidelines. Moreover, considerable attention is currently being devoted by the aviation community to support society’s efforts in the energetic transition to more sustainable solutions aiming at lower levels of greenhouse gas (GHG) emissions to oppose their consequences to the aggravation of the climate change effects. As a consequence, several initiatives are under development towards the electrification of flying vehicles as a means to achieve a more viable aviation. 

In the context of fuel-based aircraft operations, various initiatives are being carried out to keep up with the fast-paced growth of air travel demand, while providing more cost-effective solutions, both in short and long terms. One worth mentioning initiative is the incentive to replace fossil-based fuel with renewable or waste-derived Sustainable Aviation Fuel (SAF) with a lower carbon footprint \cite{b5}. Although the introduction of SAF in aviation has the greatest potential in reducing its associated GHG emissions \cite{b6},  SAF production is still far from meeting the aviation fuel demand, as it only represents 3\% of all global renewable fuels produced in 2023 \cite{b1}. For all-electric aircraft, one of the main challenges in making them viable is the limited energy density in the electric batteries systems that provide energy for the aircraft operation. In \cite{b7}, it is shown that the current state of battery development in terms of energy density is significantly far from achieving the same levels supplied by aviation fuel. 

Unfortunately, the current volume of produced SAF is insufficient to respond to the demand of fuel-based aviation and the energy density of modern batteries is unsuitable for long-haul flights. An alternative to provide more efficient energy usage in aviation is to minimize the overall trip costs by operating an aircraft in economy mode (often called ECON). This is achieved by minimizing the direct operating cost (DOC) to find optimal airspeed values. The ECON speed can be calculated by the aircraft Flight Management System (FMS) or by the airline during flight planning and preparation and it is based on the chosen ratio of time-related costs and energy-related costs, known as the cost index ($CI$).
Airlines are continuously seeking to reduce their operational costs, to increase their profit margins in a highly competitive market. Simultaneously, they are constantly challenged to exceed customer's expectations, while maintaining high levels of safety and complying with the applicable airworthiness and workforce regulations, as well as with maintenance standards. Numerous factors might affect the aircraft operation after the flight plan is approved and is under execution. One can mention (i) weather-related conditions, such as severe weather formations, precipitation, winds, lightning activities, turbulence, hail, etc.; (ii) situational conditions, such as traffic congestion, airspace occupancy, and restrictions, emergency flights that take precedence, required rerouting, etc., (iii) Air Traffic Control (ATC) coordination actions, such as in-air holdings and mandatory delays to comply with airport capacity or even (iv) the airline's decision to change the aircraft’s flight operational mode. 
To factor in the objectives of the airlines with the constraints imposed by the aircraft operation, this paper proposes a unified methodology to calculate the flight time and a constant cruise airspeed that minimize the direct operating cost of fuel-powered and all-electric aircraft in the presence of a time-varying cost index. As detailed in section \ref{sec2}, to the best of the authors' knowledge, only the references \cite{b8} -\cite{b13} available in the open literature considered a variable $CI$ once the aircraft is flying. However, none of them studied how a time-varying $CI$ affects the optimal solution of the minimization of DOC for both fuel-powered and all-electric aircraft. 
More specifically, the main contributions of this paper are:

\begin{enumerate}
\item The introduction of the cost index as a time-varying parameter in the formulation of the optimization problem to minimize DOC, which allows changes in the aircraft $CI$ imposed by the aircraft operation (input from ATC).
\item A unified approach to solve the minimization of DOC with variable $CI$ for fuel-powered and all-electric aircraft where the problem formulation considers the aircraft energy as a system state.
\item The validation of the proposed methodology by a simulated operational scenario.
\end{enumerate}

This paper is organized as follows: Section II highlights key aspects of other work in the technical literature and how they relate to the contributions of this paper. Section III presents the methodology to perform the FMS initialization and the calculation of optimal airspeed and flight time with variable $CI$. Section IV presents a simulated scenario and discussions about the results. Section V concludes the paper.

\section{Related Work}\label{sec2}
$CI$ is a trade-off parameter that balances time-related costs ($C_{t}>0$) and energy-related costs ($C_{e}>0$) in an aircraft operation and is computed by 
\begin{equation}
CI = \frac{C_{t}}{C_{e}}\label{eq1}
\end{equation}
where $C_{t}$ is defined in units of currency per units of time and it includes maintenance costs, salaries, leasing costs and equipment depreciation among other operational expenses that are time-dependent. 
The cost $C_{e}$ is associated with the cost of fuel consumed by the aircraft flying a defined route for a fuel-powered aircraft \cite{b14} or the cost of charging the electrical batteries in all-electric aircraft, and it is given in units of currency per units of energy. In this paper, $CI$ is determined in units of energy per units of time ($kJ.s^{-1}$).

The minimization of direct operating cost (DOC) has been one of the primary objectives of onboard FMS since its introduction in the early 1980s. Initial efforts to solve the minimization of DOC as an optimal control problem focused on applying Pontryagin’s Minimum Principle supported by computational simulations \cite{b15}\cite{b16}. A more recent work approached the FMS economy mode problem by combining the minimum principle with the Hamilton-Jacobi-Bellman (HJB) equation to provide a unified solution for the cruise phase as a function of $CI$ \cite{b17}. An extension of this work was proposed by the authors of \cite{b18}, where linear approximations of the ECON speed as a function of the aircraft altitude for different $CI$ values were found using the least-squares method. Article \cite{b19} proposes an open flight trajectory optimizer that includes models for fuel consumption, environmental aspects, and $CI$ in the objective function. For all-electric aircraft, article \cite{b20} proposes an optimal control framework for the FMS economy and maximum endurance modes. In \cite{b21}, the authors consider the impacts of the battery
dynamics on the overall operating cost to compute the aircraft's optimal speed based on a given $CI$. Nevertheless, once $CI$ was selected, it was assumed to be constant over the flight time. 

The authors of \cite{b8} explore the notion of a changing $CI$ based on quantifying operational factors such as passenger delay costs and ATC coordination, as well as environmental impacts, to support managing flight delay costs in pre-departure and in-flight phases. In \cite{b9}, expert decision modules using fuzzy logic are proposed to aid pilots to select a new value of $CI$ based on strategic and tactical information of the flight. In \cite{b10}, the authors combine dynamic cost indexing with waiting for passengers to minimize airline operating costs. A framework to find optimal trajectories based on minimum expected total costs is presented in \cite{b11}. A two-stage optimization procedure is applied, where first a generalized minimum DOC is obtained, based on a given $CI$, and then, a new $CI$ that minimizes the expected total costs is calculated to compensate for delays upon changes in the weather forecast. The author of \cite{b12} presents a decision-support algorithm that provides pilots with appropriate ranges for $CI$ to adjust the in-flight profile by imposing a bound in DOC. Differently from previous concepts that compute $CI$ based on deterministic optimization programming, \cite{b13} shows the application of a stochastic optimization methodology that assumes weather forecasts and payload variables with associated uncertainties and a probabilistic distribution to minimize total flight costs.

In contrast with the previous literature, this paper considers a DOC minimization problem where the $CI$ is time-varying. To the best of the authors' knowledge, this is the first time a unified approach for fuel-powered and all-electric aircraft is proposed where the $CI$ is a time-varying signal commanded by ATC. Moreover, the paper shows how a variable $CI$ affects the solution of the optimization problem as it presents the equations that allow the computation of optimal constant cruise airspeed and flight time based on step changes in the $CI$ value. The validation of the proposed methodology will be shown in section \ref{sec4}, where we present a simulation of an operational scenario inspired by a case study from \cite{b9}.

\section{Problem Statement and Solution}\label{sec3}
\subsection{Aircraft Dynamic Model and Assumptions}\label{subsec3.A}

Let us consider that $x$ describes the horizontal position of the aircraft, $v$ is its airspeed, $D$ is the magnitude of the drag force, $L$ is the magnitude of the lift force, $T$ is the magnitude of the aircraft’s thrust force, $W$ is the aircraft’s weight, and $CI$ is the aircraft cost index. Assumptions 1) to 4) are made for fuel-powered and all-electric aircraft:

\begin{enumerate}
\item The aircraft cruises in steady flight without wind, at constant altitude and with constant speed. As a consequence of this assumption, we have $W=L$ and $T=D$.
\item The flight Mach number is assumed to be below the drag divergence Mach number and wave drag
can be ignored.
\item The aircraft operates within its flight envelope.
\item The aircraft is assumed to be a fixed-wing aircraft, so the wing surface area $S$ is constant.
\end{enumerate}

For fuel-powered aircraft,  assumption 5) is considered:

\begin{enumerate}
\setcounter{enumi}{4}
\item The specific fuel consumption $S_{fc}$ is assumed to be constant for a given altitude.
\end{enumerate}

For all-electric aircraft, assumption 6) is added:

\begin{enumerate}
\setcounter{enumi}{5}
\item The battery of the all-electric aircraft is considered ideal, with neglectable internal resistance and with electric charge $Q$. It operates in cruise at a constant voltage $U$ and as a consequence, $\frac{d}{dt}(QU)=\dot Q U$.
\end{enumerate}

Based on the problem assumption 1), the aircraft cruises at constant airspeed. For any $v\neq0$, the total cruise time $t_{f}$ can then be expressed as
\begin{equation}
t_{f} = \frac{x_{f}-x_{0}}{v} = \frac{\Delta{x}}{v}\label{eq14}
\end{equation}
where $x_0$ is the starting position and $x_f$ is the final position.
It is clear that a constant cruise airspeed $v$ corresponds to a determined total flight time $t_{f}$ for a given initial position $x_{0}$ and destination $x_{f}$ as a consequence of (\ref{eq14}). We will address the aircraft scheduling in this paper assuming that the aircraft airspeed will be greater than a known value, hereafter called $v_{min}$, that corresponds to the maximum allowed flight time that does not cause significant delays at the destination.

This paper proposes a unified approach that considers the energy $E$ sourced to fuel-powered aircraft by the fuel combustion in the aircraft powerplant system (noted as $E^{fuel}$) and the energy supplied to all-electric aircraft by the electrical batteries (noted as $E^{elec}$). Thus, the time rate of change of the energy available to the aircraft from its energy source is defined as

\begin{equation}
  \dot{E} =
    \begin{cases}
      \dot{E}^{fuel}=\frac{e}{g}\dot{W}_{fuel} & \text{if fuel}\\
       \dot{E}^{elec}=\dot{Q}U & \text{if electrical}\label{eq2}
    \end{cases}       
\end{equation}
where $e$ is the fuel heating value, which is constant for a certain type of fuel and it represents the energy stored per kilogram of fuel. Typical values for jet fuel range from 40,000 $kJ/kg$ to 43,000 $kJ/kg$ \cite{b22}. The gravitational acceleration $g$ is assumed to be constant and equal to 9.81 $m.s^{-2}$, and $W_{fuel}$ is the weight of fuel available in the aircraft fuel tanks. We note that the total aircraft weight $W$ is such that the only time-varying component is $W_{fuel}$ and therefore $\dot W_{fuel}=\dot W$. In addition, $Q$ is the electrical charge available in the all-electric aircraft’s battery system.
Therefore, the dynamic model of the aircraft can be expressed as:
\begin{equation}
\dot{x}=v \label{eq3}
\end{equation}

\begin{equation}
  \dot{E} =
    \begin{cases}
      \dot{E}^{fuel}=\frac{e}{g}\dot{W} & \text{if fuel}\\
       \dot{E}^{elec}=\dot{Q}U & \text{if electrical}\label{eq5}
    \end{cases}       
\end{equation}

The magnitude of the drag force is
\begin{equation}
D=\frac{1}{2}\rho S C_{D}v^{2} \label{eq6}
\end{equation}
where $C_{D}$ is the drag coefficient. As per assumption 2), the aircraft operates below the drag divergence Mach number, and assuming that it follows a drag polar curve, one can define
\begin{equation}
C_{D}=C_{D,0}+C_{D,2}C_{L} ^{2} \label{eq7}
\end{equation}
where $C_{D,0}$ is the parasitic drag coefficient at zero-lift, $C_{D,2}$ is the drag coefficient induced due to lift and $C_{L}$ is the lift coefficient, which can be determined by
\begin{equation}
C_{L}=\frac{2L}{\rho S v^{2}} \label{eq8}
\end{equation}

Replacing (\ref{eq7}) and (\ref{eq8}) in (\ref{eq6}) and considering $L = W$ as per assumption 1), then the drag force can be expressed as
\begin{equation}
D=\frac{1}{2}\rho S C_{D,0}v^{2} + \frac{2C_{D,2}W^{2}}{\rho S v^{2}}\label{eq9}
\end{equation}
\subsection{Problem Formulation}\label{subsec3.B}
The direct operating cost (DOC) of an aircraft in cruise is
\begin{equation}
DOC=\int_{0}^{t_{f}} (C_{t}-C_{e}\dot{E}) \,dt\label{eq10}
\end{equation}
where $t_{f}$ is the total cruise time and the minus sign corresponds to energy depletion. This notion can be expanded to a flight composed of several cruise segments, where $t_{f}$ indicates the total flight time in each of these segments. Assuming that the cost of energy $C_{e}$ is constant, one can divide the cost function (\ref{eq10}) by $C_{e}$, yielding
\begin{equation}
J = \frac{DOC}{C_{e}} = \int_{0}^{t_{f}} (CI -\dot{E}) \,dt\label{eq11}
\end{equation}

As discussed in Section \ref{sec2}, most of the related work in the literature considers the cost index as a constant value throughout the flight or between waypoints. However, in this paper, we explore the properties of a variable cost index function modeled as a first-order filter. 

\begin{figure}[htbp]
\centerline{\includegraphics[scale=0.25]{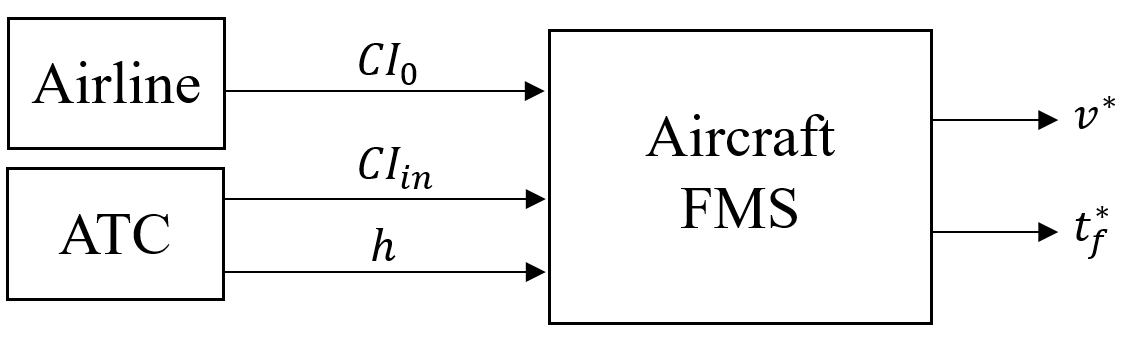}}
\caption{Block diagram of minimum DOC with ATC input}
\label{fig1}
\end{figure}

The block diagram in Figure \ref{fig1} summarizes the problem, where $CI(t_{0})=CI_{0}$, corresponds to the initial condition of the cost index, selected by the airline for the FMS initialization when preparing the aircraft flight plan and schedule. The ATC determines the flight level $h$ for the aircraft cruise. As previously discussed, changes in the aircraft's $CI$ are expected throughout the flight from the ATC to adjust the air traffic flow in a certain airspace. The magnitude of the step change in the aircraft $CI$, noted as $CI_{in}$ is a result of multiple factors that depend on environmental, situational, or operational conditions. There are some methodologies available in the open literature, such as in \cite{b9} and \cite{b11}, that could be used by the ATC to compute the $CI_{in}$. Therefore, we assume that $CI_{in}$ is also provided by the ATC to the aircraft FMS along with the cruise flight level $h$. The result of the optimization problem will be the optimal aircraft airspeed $v^{*}$ and flight time $t_{f}^{*}$.

In this context, the minimization of DOC for an aircraft in cruise with constant airspeed and variable cost index can be formulated as an optimal control problem as shown below
 
\begin{equation}
\begin{aligned}
J^{*} = \min_{v,t_{f}} \quad & \int_{0}^{t_{f}} (CI -\dot{E}) \,dt \label{eq13}\\
\textrm{s.t.} \quad & \dot{x}=v\\
  &\dot{E} =
    \begin{cases}
      \dot{E}^{fuel}=\frac{e}{g}\dot{W} & \text{if fuel}\\
       \dot{E}^{elec}=\dot{Q}U & \text{if electrical}
    \end{cases}      \\
  &\tau\dot{CI}=-CI+CI_{in}\\
  &D=\frac{1}{2}\rho S C_{D,0}v^{2} + \frac{2C_{D,2}W^{2}}{\rho S v^{2}}\\
  &CI(0)=CI_{0}\\
  &E(0)=E_{0}\\
  &x(0)=x_{0}, x(t_{f})=x_{f}\\
  &v>v_{min}
\end{aligned}
\end{equation}
where $J^{*}$ is the minimum DOC achieved for the minimizers of (\ref{eq13}), which are the optimal airspeed $v^{*}$ and the optimal flight time $t_{f}^{*}$.

\textit{Remark:} As explained in section \ref{subsec3.A}, this paper addresses aircraft scheduling  by introducing a lower bound on the aircraft's airspeed, to prevent significant delays at the destination.

\subsection{Problem Solution}\label{subsec3.C}
The solution of the equation in $\dot{CI}$ in (\ref{eq13}), for the initial condition $CI(t_{0})=CI_{0}$ and input $CI_{in}$, is given by
\begin{equation}
CI(t)=e^{-\frac{t}{\tau}}(CI_{0}-CI_{in})+CI_{in} \label{eq17}
\end{equation}
where $\tau$ is the time constant of the first-order filter and indicates the convergence rate of the $CI$ to reach the commanded value $CI_{in}$. Considering the $CI$ now as a function of time, one can rewrite the total cost function $J$ from the optimization problem (\ref{eq13}) using the result from (\ref{eq17}) and (\ref{eq14}) as
\begin{equation}
J=\tau(CI_{0}-CI_{in})(1-e^{-\frac{\Delta{x}}{\tau v}})+\frac{CI_{in}\Delta{x}}{v}+E_{0} -E_{f}\label{eq18}
\end{equation}
where $E_{f}$ is the final value of the aircraft energy. Applying the necessary condition for optimality on (\ref{eq18}) yields

\begin{equation}
\frac{\partial J}{\partial v} = -\frac{(CI_{0}-CI_{in})\Delta{x}}{v^{2}}e^{-\frac{\Delta{x}}{\tau v}} -CI_{in}\frac{\Delta{x}}{v^{2}}-\frac{\partial E_{f}}{\partial v}=0\label{eq19}
\end{equation}

The optimal cruise airspeed $v^{*}$ for a variable $CI$ is the solution of (\ref{eq19}), for any finite $\tau>0$, $v>0$ and $x_{f}>x_{0}$. The partial derivative of the final energy with respect to the aircraft's airspeed depends on the type of aircraft and will be discussed in section \ref{subsecD}.

\subsubsection{FMS Initialization}\label{subsubsec1} 
If no ATC input is received throughout the flight, the aircraft should operate with the fixed value of $CI=CI_{0}$, which is the $CI$ value defined by the airline based on its strategy. In this case, a particular solution can be derived for the FMS initialization by making $\tau=\infty$ in (\ref{eq19}), which results in
\begin{equation}
-CI_{0}\frac{\Delta{x}}{v_{0}^{2}}-\frac{\partial E_{f}}{\partial v_{0}}=0\label{eq16}
\end{equation}
where $v^{*}_{0}$ is the optimal airspeed computed for the FMS initialization. The optimal flight time $t_{f_{0}}^{*}$ can be computed using (\ref{eq14}) with $v=v^{*}_{0}$.

\subsection {Application to fuel-powered and all-electric aircraft}\label{subsecD}
As mentioned in section \ref{subsec3.C}, to compute the optimal cruise airspeed and flight time for the FMS initialization and for cruise operations with ATC input, one needs first to compute the term $\frac{\partial E_{f}}{\partial v}$, which depends on the aircraft’s energy supply, i.e. if the aircraft is fuel-powered or all-electric. This section shows how to apply equations (\ref{eq14}), (\ref{eq19}) and (\ref{eq16}) to compute the optimal airspeed and flight time for fuel-powered and all-electric aircraft. 

\subsubsection{Fuel-powered aircraft} \label{subsubsecD1}
For a fuel-powered aircraft, from (\ref{eq2}), we can establish a relationship between the final weight $W(t_{f})=W_{f}$ and the final fuel energy $E_{f}^{fuel}$ as
\begin{equation}
E_{f}^{fuel}=\frac{e}{g} W_{f}\label{eq20}
\end{equation}

The time rate of change of the aircraft weight in steady flight is expressed by
\begin{equation}
\dot{W}=-S_{fc}D=-S_{fc}\Biggl(\frac{1}{2}\rho S C_{D,0}v^{2} + \frac{2C_{D,2}W^{2}}{\rho S v^{2}}\Biggl)\label{eq21}
\end{equation}

The solution of the separable differential equation (\ref{eq21}) with initial condition $W_{0}$ is
\begin{equation}
W_{f}=k_{2}v^{2}tan\Biggl[-\frac{\Delta{x}}{k_{1}v}+arctan\Biggl(\frac{W_{0}}{k_{2}v^{2}}\Biggl)\Biggl]\label{eq22}
\end{equation}
where the constants $k_{1}$ and $k_{2}$ are defined as
\begin{equation}
k_{1}=\frac{1}{S_{fc}\sqrt{C_{D,0}C_{D,2}}}\label{eq23}
\end{equation}
\begin{equation}
k_{2}=\frac{\rho S}{2}\sqrt{\frac{C_{D,0}}{C_{D,2}}}\label{eq24}
\end{equation}

Replacing (\ref{eq22}) in (\ref{eq20}) yields
\begin{equation}
E_{f}^{fuel}=\frac{e}{g} k_{2}v^{2}tan\Biggl[-\frac{\Delta{x}}{k_{1}v}+arctan\Biggl(\frac{W_{0}}{k_{2}v^{2}}\Biggl)\Biggl]\label{eq25}
\end{equation}

Based on (\ref{eq25}) we can compute $\frac{\partial E_{f}}{\partial v}$, which for a fuel-powered aircraft becomes
\begin{equation}
\begin{aligned}
\frac{\partial E_{f}^{fuel}}{\partial v}=2\frac{e}{g}k_{2}v tan\Biggl[-\frac{\Delta{x}}{k_{1}v}+arctan\Biggl(\frac{W_{0}}{k_{2}v^{2}}\Biggl)\Biggl]+\\
\frac{e}{g}\Biggl(\frac{k_{2}\Delta{x}}{k_{1}}-\frac{2v^{3}W_{0}}{v^{4}+\frac{W_{0}^{2}}{k_{2}^{2}}}\Biggl)sec^{3}\Biggl[-\frac{\Delta{x}}{k_{1}v}+arctan\Biggl(\frac{W_{0}}{k_{2}v^{2}}\Biggl)\Biggl]\label{eq26}
\end{aligned}
\end{equation}

Replacing (\ref{eq25}) in (\ref{eq18}), we obtain the total cost function for the FMS in cruise with ATC input for a fuel-powered aircraft. The result from (\ref{eq26}) can be applied to (\ref{eq19}) and (\ref{eq16}) to compute the optimal cruise airspeed for the aircraft cruise with ATC input and for the FMS initialization, respectively, for fuel-powered aircraft. The optimal flight time can be calculated by (\ref{eq14}) based on the optimal cruise airspeed.

\subsubsection{All-electric aircraft}\label{subsubsecD2}
For an all-electric aircraft, we know from (\ref{eq2}) that the total energy stored in the aircraft’s battery system $E^{elec}$ depends on the electrical charge of the batteries and the battery voltage. As per assumption 6), for a constant voltage $U$, the final energy $E_{f}^{elec}$ available to the aircraft is
\begin{equation}
E_{f}^{elec}=Q_{f}U\label{eq27}
\end{equation}
where $Q(t_{f})=Q_{f}$ is the final battery charge. From the definition of efficiency in the conversion of the electrical power to mechanical power, the electrical current $i$ supplied by the aircraft battery, which is equal to the time rate of change in the battery’s charge $\dot{Q}$, can be expressed as
\begin{equation}
i=-\dot{Q}=-\frac{Tv}{U \eta}\label{eq28}
\end{equation}

For steady flight, $T = D$, so (\ref{eq28}) and (\ref{eq9}) lead to 
\begin{equation}
\int_{Q{0}}^{Q_{f}}\,dQ = -\Biggl(\frac{\rho S C_{D,0}v^{3}}{2U \eta} + \frac{2C_{D,2}W^{2}}{U \eta \rho S v}\Biggl) \int_{0}^{t_{f}}\,dt\label{eq29}
\end{equation}

Using the result of (\ref{eq14}), the solution of (\ref{eq29}) is
\begin{equation}
Q_{f} = Q_{0} - \frac{\Delta{x}}{U \eta}\Biggl(\frac{1}{2}\rho S C_{D,0}v^{2} + \frac{2C_{D,2}W^{2}}{\rho S v^{2}}         \Biggl)\label{eq30}
\end{equation}

Replacing (\ref{eq30}) in (\ref{eq27}), we obtain
\begin{equation}
E^{elec}_{f} = Q_{0}U - \frac{\Delta{x}}{\eta}\Biggl(\frac{1}{2}\rho S C_{D,0}v^{2} + \frac{2C_{D,2}W^{2}}{\rho S v^{2}}         \Biggl)\label{eq31}
\end{equation}

From (\ref{eq31}), we can now compute $\frac{\partial E_{f}}{\partial v}$, which for an all-electric aircraft becomes
\begin{equation}
\frac{\partial E_{f}^{elec}}{\partial v} = - \frac{\Delta{x}}{\eta}\Biggl(\rho S C_{D,0}v -\frac{4C_{D,2}W^{2}}{\rho S v^{3}}\Biggl)\label{eq32}
\end{equation}

With the result (\ref{eq31}), one can obtain the total cost function $J$ for the FMS operating in cruise with an ATC input as per (\ref{eq18}). Replacing (\ref{eq32}) in (\ref{eq19}) and (\ref{eq16}), one can compute the optimal airspeed for the aircraft cruise operation under ATC input and for the FMS initialization, respectively, for an all-electric aircraft.

\subsection {Sufficient Condition for optimality}\label{subsecE}

To confirm that the optimal airspeed $v^*$ is a minimizer of the total cost function $J$ in (\ref{eq18}), the sufficient condition for optimality (\ref{eq33}) shall be satisfied.

\begin{equation}
\begin {aligned}
\frac{\partial^{2} J}{\partial v^{2}} = \frac{(CI_{0}-CI_{in})\Delta{x}e^{-\frac{\Delta{x}}{\tau v}}}{v^{4}} \Biggl(2v-\frac{\Delta{x}}{\tau}\Biggl) + \\ \frac{2CI_{in}\Delta{x}}{v^{3}} - \frac{\partial^2 E_{f}}{\partial v^2} > 0 \label{eq33}
\end {aligned}
\end{equation}

For all-electric aircraft, the second derivative of the final energy with respect to the airspeed is given by

\begin{equation}
\frac{\partial^2 E_{f}^{elec}}{\partial v^2} = -\frac{\Delta{x}}{\eta}\Biggl(\rho S C_{D,0} + \frac{12C_{D,2}W^2}{\rho S {v}^4}\Biggl) \label{eq34}
\end{equation}

For fuel-powered aircraft, the second derivative of the final energy with respect to the airspeed is computed as

\begin{equation}
\begin{aligned}
\frac{\partial^2 E_{f}^{fuel}}{\partial v^2} = 2 \frac{e}{g}k_{2} \Biggl[tan(\alpha) - k_{3} v sec^2(\alpha) +     
\\ \frac{8eW_{0}v^3}{[v^4 + (\frac{W_{0}}{k_{2}})^2]^2}sec^3(\alpha) + \\
3k_{3}\Biggl[\frac{ek_{2}\Delta{x}}{gk_{1}} -\frac{2eW_{0}}{g\Bigl(v^4 +(\frac{W_{0}}{k_{2}})^2 \Bigl)}\Biggl]sec^3(\alpha)tan(\alpha)
\label{eq35}
\end{aligned}
\end{equation}

where

\begin{equation}
\alpha = -\frac{\Delta{x}}{k_1 v} + arctan\Bigl(\frac{W_{0}}{k_2 v^2}\Bigl) \label{eq36}
\end{equation}

\begin{equation}
    k_{3} = -\frac{\Delta{x}}{k_{1} v^2} + \frac{2(\frac{W_{0}}{k_{2}})v}{v^4 + (\frac{W_{0}}{k_{2}})^2} \label{eq37}
\end{equation}

\begin{figure}[htbp]
\centerline{\includegraphics[scale=0.35]{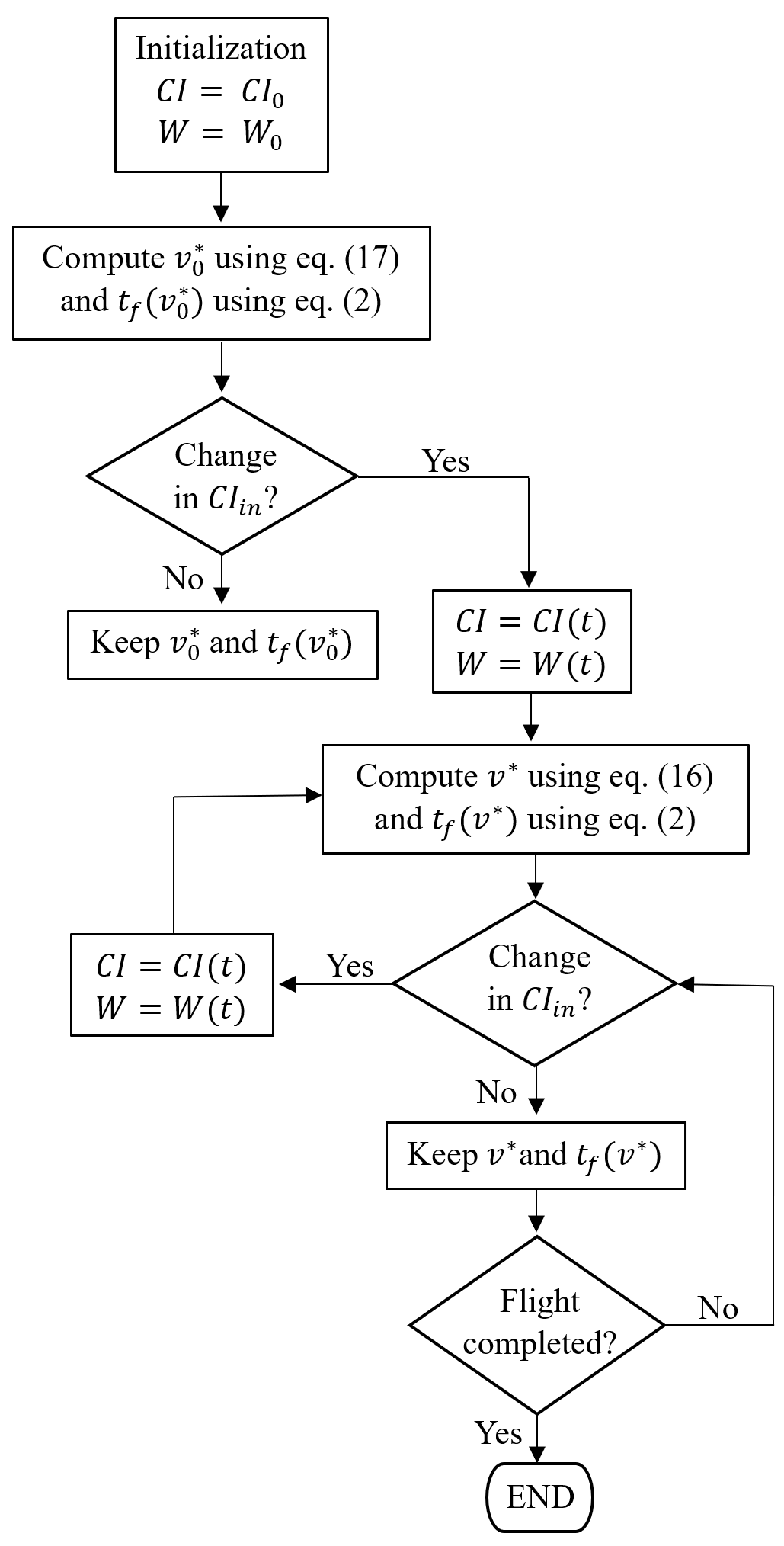}}
\caption{Optimal airspeed and flight time calculation flowchart}
\label{fig2}
\end{figure}

The flowchart shown in Figure \ref{fig2} describes how the equations derived in this paper can be used in an actual operational scenario. The initial value of the optimal airspeed is computed using (\ref{eq16}). Once the aircraft cruise is ongoing, any changes in the aircraft $CI$ by the ATC will make the FMS calculate a new optimal airspeed as per (\ref{eq19}) until the flight is completed. The flight time is computed by (\ref{eq14}). 
\\
\section{Results and Discussions}\label{sec4}
\subsection{Simulation Parameters} \label{subsec4a}
The simulations presented herein were performed in MATLAB installed on a laptop equipped with 16 GB of RAM and an $11^{th}$ Gen Intel(R) Core(TM) i5-1135G7 2.40GHz CPU. To simulate a fuel-powered aircraft, we used data from a Gulfstream-IV(G-IV) business jet \cite{b17}, \cite{b23}, while for an all-electric aircraft, data from a Yuneec International E430 two-seater aircraft model \cite{b24} is used as reference for the simulated scenario, as presented in Table \ref{table1}.

\begin{table} [hbt]
  \begin{threeparttable}
   \caption{Simulation Parameters}\label{table1}
   \centering
   \begin{tabular}{|c c c |}
   \hline
     \textbf{Parameters} & \textbf{Fuel-powered} & \textbf{All-electric}\\
     \hline
         Wing surface area $S$ $(m^{2})$ & 88.26 & 11.37 \\ 
         Aircraft Mass $(kg)$ & 10000\tnote{1} & 472\tnote{2} \\
         Zero-lift drag coefficient $C_{D,0}$ & 0.015 & 0.035 \\
         Induced drag coefficient $C_{D,2}$ & 0.08 & 0.009\\
         Maximum cruise airspeed $v_{max}$ $(km/h)$ & 890 & 161\\
         Specific Fuel Consumption $S_{fc}$ $(kg/Ns)$ & 1.92x10$^{-5}$ & N/A \\ 
         Battery Output Voltage $U$ $(V)$ & N/A & 133.2\\
         Electrical system efficiency $\eta$ & N/A & 0.7\\ [1ex] 
     \hline
     \end{tabular}
     \begin{tablenotes}
       \item [1] Initial fuel mass.
       \item [2] Aircraft maximum take-off mass, which is constant throughout the flight.
     \end{tablenotes}
  \end{threeparttable}
\end{table}

\subsection{Simulated flight scenario}\label{subsec4b}
\begin{figure}[htbp]
\centerline{\includegraphics[scale=0.28]{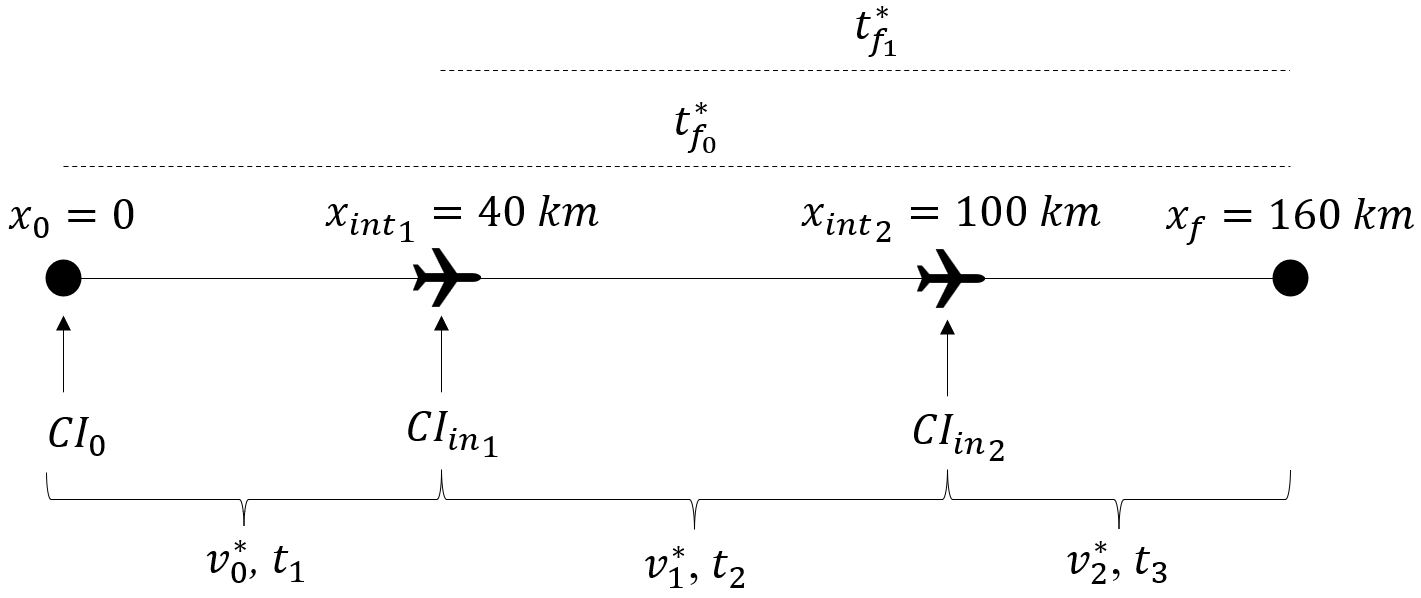}}
\caption{Flight scenario}
\label{fig3}
\end{figure}

A simulation of a flight scenario where the optimal cruise airspeed and flight time are found for a fuel-powered and an all-electric aircraft is presented in this section. The optimal aircraft airspeed values were computed in MATLAB by the \textit{fzero} function using (\ref{eq19}) and (\ref{eq16}) and the flight time was found by solving (\ref{eq14}). In the simulated scenario, the ATC imposed two changes in the aircraft airspeed to conform the aircraft operation to the current air traffic conditions in the area where it is flying in, so the aircraft airspeed shall be adjusted, based on a $CI$ given by the ATC. Let us consider that both a fuel-powered and an all-electric aircraft with characteristics presented in Table \ref{table1} are flying from the initial position $x_{0}=0$ (considering initial time $t_{0}=0s$) to the destination position $x_{f}=160 km$ and with initial cost index $CI_{0}=0.1CI_{max}$, assuming $0 \leq CI \leq CI_{max}$, where $CI_{max}$ corresponds to the maximum value of $CI$, with the aircraft operating within its envelope, as per assumption 3). The optimal cruise airspeed $v_{0}^{*}$ was computed using (\ref{eq16}) with $CI_{0}=0.1CI_{max}$. The flight time, which in this case is the scheduled flight time $t_{f_{0}}^{*}$ is found using (\ref{eq14}). When the aircraft is at the first intermediate position $x_{int_{1}}=40 km$, after time $t_{1}$ has elapsed, an ATC input $CI_{in_{1}}=0.2CI_{max}$ is received, so the aircraft could recover from a delay in its departure. The adjusted optimal airspeed in the second flight segment $v^{*}_{1}$ and the flight time $t_{f_{1}}^{*}$ will be then computed using (\ref{eq19}) and (\ref{eq14}), respectively, with $x_{0}=x_{int_{1}}$. When the aircraft is at the second intermediate position $x_{int_{2}}$, which is 100 $km$ distant from the origin waypoint, after time $t_{2}$ has passed, a new ATC input $CI_{in_{2}}=0.15CI_{max}$ is received, and the aircraft is required to decrease its airspeed. The new optimal airspeed $v^{*}_{2}$ and flight time $t_{3}$ in the third flight segment were computed using (\ref{eq19}) and (\ref{eq14}), respectively, now with $x_{0}=x_{int_{2}}$. Figure \ref{fig3} summarizes the described flight scenario.

This flight scenario is inspired by the case study presented in \cite{b9}, where the aircraft $CI$ is adjusted in flight. However, in the case study from \cite{b9}, the aircraft flew from Madrid to Moscow, which is a mission appropriate for fuel-powered aircraft that are capable of long-haul flights, as opposed to all-electric aircraft that typically operate in short-range flights. Therefore, the distance between the origin and the destination in the simulation scenario herein proposed was changed to create an environment where both a fuel-powered and an all-electric aircraft can operate, while keeping similar operational restrictions as presented in \cite{b9}. The case study presented in \cite{b9} exclusively provides the new Estimated Time of Arrival (ETA) at the destination. In contrast, this paper computes the optimal airspeed based on changes in $CI$ in cruise imposed by the ATC and shows how the aircraft's airspeed reaches the optimal value. Moreover, we show how the changes in $CI$ impact the aircraft ETA at the destination, as the optimal flight time is also computed by the method proposed in our paper.

The flight level $h$ is also determined by the ATC and is suitable for each type of aircraft. For the fuel-powered aircraft, we assumed a constant altitude of 10km above sea level, which corresponds to an air density $\rho = 0.4135$ $kg/m^{3}$ and $S_{fc}$ as defined in Table \ref{table1}. For the all-electric aircraft, the altitude considered in the simulation was 1km above sea level, where the air density is considered as $\rho = 1.112$ $kg/m^{3}$.

\subsubsection {Time constant $\tau$}\label{subsubsec4B1}
\begin{figure}[hbt!]
\centerline{\includegraphics[scale=0.45]{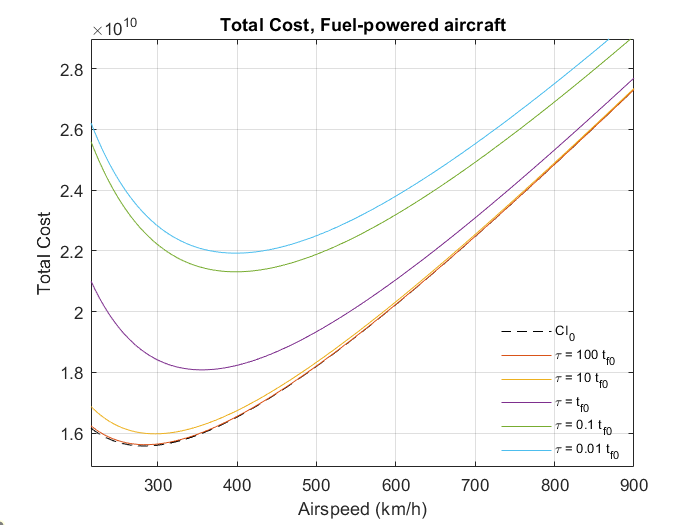}}
\caption{Total Cost as a function of the fuel-powered aircraft’s airspeed}
\label{fig5}
\end{figure}

\begin{figure}[hbt!]
\centerline{\includegraphics[scale=0.45]{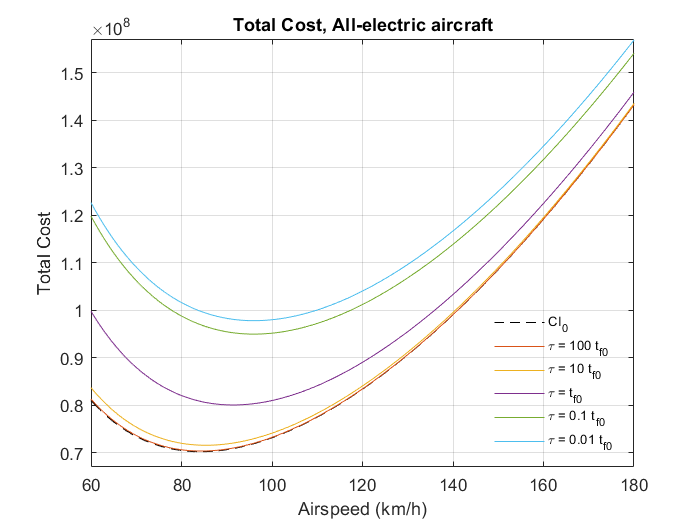}}
\caption{Total Cost as a function of the all-electric aircraft’s airspeed}
\label{fig6}
\end{figure}

In this section, we show the effect of the time constant $\tau$ on the behavior of the variable cost index in the total cost function as per (\ref{eq18}). The parameter $\tau$ expresses the decay of the $CI$ function and how fast it converges to $CI_{in}$. Figure \ref{fig5} and Figure \ref{fig6} show the total cost as a function of the aircraft's airspeed for different values of the time constant $\tau$, for a fuel-powered and an all-electric aircraft, respectively. As a reference, the dashed line represents the total cost function for an aircraft operating with a constant $CI=CI_{0}$. In this example, $CI_{in}>CI_{0}$. The optimal cruise  airspeed for larger values of $\tau$ is smaller than the optimal cruise airspeed computed for smaller values of $\tau$. As a consequence, the total energy consumption is also smaller for aircraft operating with higher values of $\tau$. As the ATC imposed an operational restriction on the ongoing flight, the aircraft shall adjust its airspeed to comply with the ATC input. Smaller values of $\tau$ cause $CI$ to converge faster to $CI_{in}$, reducing the risks of non-compliance with the ATC requirement. Based on the observed behavior of $CI$ for different values of $\tau$, a value of $\tau = 0.01t_{f_{0}}^{*}$, was chosen as the first-order filter parameter considered in this paper.


\subsubsection {Cost index and airspeed}\label{subsubsec4B2}
\begin{figure}[hbt!]
\centerline{\includegraphics[scale=0.45]{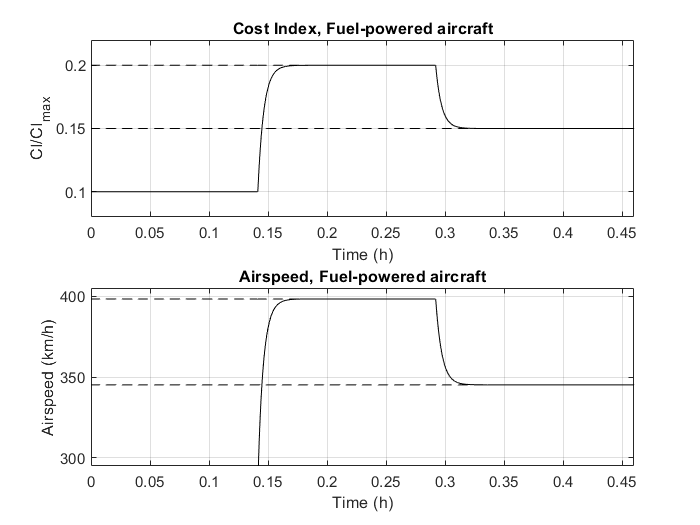}}
\caption{Cost index (top) and airspeed (bottom) as a function of flight time, fuel-powered aircraft}
\label{fig7}
\end{figure}

\begin{figure}[hbt!]
\centerline{\includegraphics[scale=0.45]{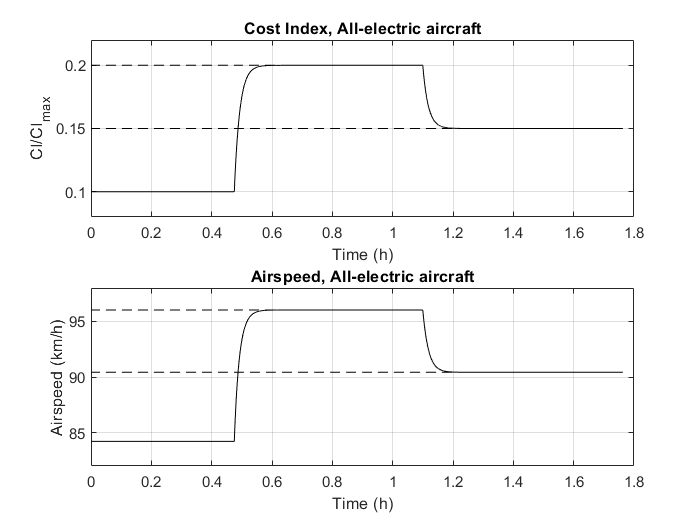}}
\caption{Cost index (top) and airspeed (bottom) as a function of flight time, all-electric aircraft}
\label{fig8}
\end{figure}

Figure \ref{fig7} and Figure \ref{fig8} depict the cost index (top) and the aircraft's airspeed (bottom) as a function of the total flight time, for the fuel-powered and the all-electric aircraft considered in this paper, respectively. The value of the cost index converges to $CI_{in_{1}}$ and $CI_{in_{2}}$ with a time constant of $\tau$. As previously stated, the parameter $\tau$ is chosen in such a way that the $CI$ converges fast to the commanded value $CI_{in}$ and the aircraft's airspeed also rapidly transitions to the optimal solutions that accommodate the ATC inputs as per (\ref{eq19}).


\subsubsection {Energy consumption}\label{subsubsec4B3}
\begin{figure}[hbt!]
\centerline{\includegraphics[scale=0.45]{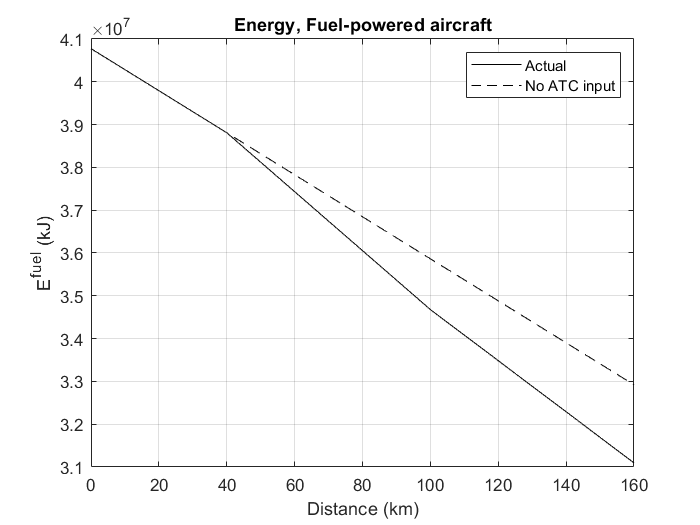}}
\caption{Available energy as a function of distance travelled, fuel-powered aircraft}
\label{fig9}
\end{figure}

\begin{figure}[hbt!]
\centerline{\includegraphics[scale=0.45]{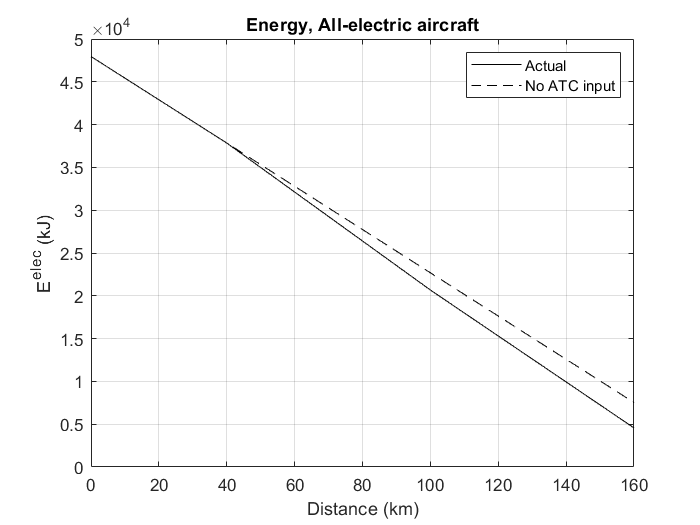}}
\caption{Available energy as a function of distance travelled, all-electric aircraft}
\label{fig10}
\end{figure}

Figure \ref{fig9} and Figure \ref{fig10} show the energy available to the aircraft as a function of the distance traveled, for a fuel-powered and an all-electric aircraft, respectively. The dashed line represents the available energy if the aircraft operated as per its original schedule, with no ATC input. However, to comply with the ATC inputs, the aircraft's airspeed was increased in the second segment and decreased in the third segment, but both revised airspeed values were higher than the airspeed that corresponds to the FMS initialization cost index. These changes resulted in a higher total energy consumption, represented by a smaller value in the final available energy. 

Applying the proposed methodology for the simulated scenario, the optimal airspeed and flight time for the fuel-powered and the all-electric aircraft were determined. Due to the changes in the airspeed to comply with the ATC input, the arrival time at the destination was also revised and the difference between the original scheduled arrival time and the actual arrival time at the destination is presented in Table \ref{tab2} as $\Delta t_{arrival}$ along with the other results. 
\begin{table} [hbt]
\begin{center}
\caption{Simulation Results}\label{tab2}
\begin{tabular}{|c c c|} 
 \hline
 \textbf{Parameter} & \textbf{Fuel-powered} & \textbf{All-electric}\\ 
 \hline\hline
 $v_{0}^{*}$ $(km/h)$ & 283.03 & 84.21 \\ 
 $t_{f_{0}}^{*}$ & 33min55s & 1h54min \\
 $t_{1}$ & 08min28s & 28min30s \\
 $v^{*}_{1}$ $(km/h)$ & 398.24 & 96.02 \\
 $t_{f_{1}}^{*}$ & 18min46s & 1h14min59s \\
 $t_{2}$ & 09min02s & 37min29s \\
 $v^{*}_{2}$ $(km/h)$ & 345.16 & 90.42 \\
 $t_{3}$ & 10min25s & 39min49s \\
 $\Delta t_{arrival}$ & -6min & -8min12s\\
 \hline
\end{tabular}
\end{center}
\end{table}

\section{Conclusions}\label{sec5}
This paper introduces a novel unified approach of considering a time-varying cost index in the minimization of DOC, to compute the optimal constant cruise airspeed and flight time for fuel-powered and all-electric aircraft. To the best of the authors’ knowledge, this is the first time a unified approach for both fuel-powered and all-electric aircraft has been proposed where $CI$ is a time-varying signal. The proposed methodology was validated by a simulated flight scenario.
In this scenario the inputs from the ATC were received during flight and the aircraft was required to adjust its optimal airspeed, flight time, and total energy consumption to comply with the operational restrictions imposed by the ATC.
The optimal values of airspeed, flight time and energy consumption were computed for both a fuel-powered and an all-electric aircraft, thus enabling applications of the proposed approach to future air mobility all-electric vehicles.

\end{document}